\documentstyle[aps,epsf,floats]{revtex}

\newcommand{\Qcor}{{\raisebox{0.2ex}{$\not$}}Q}

\ifx\epsffile\undefined
\def\insertfig#1{}
\else
\def\insertfig#1{{\baselineskip=4pt
\centerline{\epsfxsize=\hsize\epsffile{#1}}}}\fi

\begin{document}

\twocolumn[       
{\tighten
\preprint{\vbox{
\hbox{}
\hbox{}
}}
\draft
\title{\Large \bf  TWO-PION EXCHANGE NUCLEON-NUCLEON POTENTIAL:
MODEL INDEPENDENT FEATURES}
\author{Manoel R. Robilotta\footnotemark  and Carlos A. da Rocha\footnotemark } 
\address{University of Washington, Department of Physics, Box 351560,  
Seattle, Washington 98195-1560}
\bigskip
\bigskip
\date{November 1996}
\maketitle
\widetext
\vskip-2.0in
\rightline{\fbox{
\hbox{DOE/ER/41014-02-N97}
}}
%
\vskip 1.5in

\begin{abstract}
A chiral pion-nucleon amplitude supplemented by the HJS
subthreshold coefficients is used to calculate the 
the long range part of the two-pion exchange nucleon-nucleon
potential. In our expressions the HJS coefficients
factor out, allowing a clear identification of the origin 
of the various contributions. A discussion of the configuration space 
behaviour of the loop integrals that determine the potential
is presented, with emphasis on cancellations associated with 
chiral symmetry. The profile function for the scalar-isoscalar 
component of the potential is produced and shown
to disagree with those of
several semi-phenomenological potentials. 
\end{abstract}

}

] 
\narrowtext

\footnotetext{${}^*$Permanent address: Instituto de F\'{\i}sica, Universidade
de S\~{a}o Paulo, C.P. 66318, 05389-970 S\~{a}o Paulo, SP, Brazil.
Electronic address: robilotta@if.usp.br}
\footnotetext{${}^\dagger$Fellow from CNPq Brazilian Agency. Electronic address:
carocha@phys.washington.edu}


\section{INTRODUCTION}

Nowadays there are several semi-phenomenological NN potentials that may
be considered as realistic because they provide good 
descriptions of cross sections, scattering amplitudes and phase shifts.
Nevertheless, it is possible to notice important discrepancies when one 
compares directly their configuration space profile functions.
Of course, this situation
is consistent with the venerable inverse scattering problem, whereby 
there are always many potentials that can explain a given
set of observables. Therefore one must look elsewhere in order to
assess the merits of the various possible models.

In the case of NN interactions, there is a rather rich relationship between
the potential and observables, involving several spin and isospin
channels and different spatial regions. On the other hand, as
all modern models represent the long range interaction by means of
the one pion exchange potential (OPEP), one must go to
inner regions in order to unravel the discrepancies among
 the various approaches.

Models vary widely in the way they treat the non-OPEP part of the
interaction and, in the literature, one finds potentials constructed
by means of dispersion relations, field theory or just based on common
sense guesses. In all cases, parameters are used which either reflect
knowledge about other physical processes or are adjusted ad hoc. This
leaves a wide space for personal whim and indicates the need of 
information with little model dependence about the inner part of the
nuclear force. In the case of NN interactions,
the complexity of the relevant physical processes increases very
rapidly as the internucleon distance decreases and hence the best
process for yielding information with little model dependence is the
tail of the two-pion exchange potential ($TPEP$).

This problem has a long history. More than thirty years ago, Cottingham
and Vinh Mau began a research program based on the idea that the 
$TPEP$ is related to the pion--nucleon ($\pi N$) amplitude~\cite{Vin63}.
It lead to the construction of the Paris potential~\cite{Cott73,Laco80}, where 
the intermediate part of the force is obtained from empirical $\pi N$ 
information treated by means of dispersion relations. This procedure 
minimizes the number of unnecessary hypotheses and hence yields results
which can be considered as model independent. Another important contribution
was made by Brown and Durso~\cite{Brow71}  who stressed, in the early 
seventies, that chiral symmetry has a main role in the description
of the intermediate $\pi N$ amplitude.  

In the last four years the interest in applications
of chiral symmetry to nuclear problems was renewed
and several authors have reconsidered the construction of
the $TPEP$. At first, only systems containing pions and nucleons were
studied, by means of non-linear lagrangians based on either PS or PV
pion-nucleon couplings~\cite{Ordo92,Cele92,Fria94,Rocha94,Birse94}.
Nowadays, the evaluation of this part of the potential in 
the framework of chiral symmetry has no important
ambiguities and is quite well understood.
This minimal $TPEP$ fulfills the expectations from chiral
symmetry and, in particular, reproduces at the nuclear level the
well known cancellations of the intermediate $\,\pi N\,$ 
amplitude~\cite{Ballot96a,Ballot96b}. On the other hand, it fails to 
yield the qualitative features of the medium range 
scalar-isoscalar NN attraction~\cite{Rocha94,Rocha95}.
This happens because a system containing just pions and nucleons cannot
explain the experimental $\pi N$ scattering
data~\cite{HohI} and one needs other degrees of freedom, especially
those associated with the delta and the $\pi N$ $\sigma$-term.
The former possibility was considered by
Ord\'{o}\~nez, Ray and Van Kolck~\cite{Ordo94,Ordo96}, and shown to improve
the predictive power of chiral cancellations but, in their work 
they did not regard closely the experimental features of the 
intermediate $\pi N$ amplitude.

Empirical information concerning the
intermediate $\pi N$ process may be introduced into the $TPEP$
in a model independent way, with the
help of the H\"ohler, Jacob and Strauss (HJS) subthreshold 
coefficients~\cite{HohI,Hoh72}.  This kind of approach has already been 
extensively adopted in other problems. For instance, 
Tarrach and M.Ericson used it in their study of the 
relationship between nucleon polarizability and nuclear Van der Waals 
forces~\cite{Tarra78}. In the case of
three-body forces, it was employed in the construction of both
model independent and model dependent two-pion exchange 
potentials~\cite{Coon75,Rob83,Ueda84}.
Using the same strategy, we have recently 
shown that the knowledge of the 
$\pi N$ amplitude, constrained by both chiral symmetry and 
experimental information in the form of the HJS coefficients,
provides a unambiguous and model independent
determination of the long range part of the two-pion
exchange NN potential~\cite{Rob96}. There we restricted ourselves to 
the general formulation of the problem and to the identification of 
the leading scalar-isoscalar potential. 
In the present work we explore the numerical consequences
of the expressions derived
in that paper and compare them with some existing potentials.

Our presentation is divided as follows: in Sec. II, we briefly
summarize the derivation of the potential and recollect the main
formulae for the sake of self-consistency, 
leaving details to Appendices A, B, and C. In Sec. III we discuss the
main features of the loop integrals that determine the potential,
emphasizing in the approximations associated with chiral symmetry.
In Sec. IV we relate our theoretical expressions with those of other
authors and in Sec. V, results are compared with existing
phenomenological potentials. Finally, in Sec. VI we present our
conclusions.


\section{TWO-PION EXCHANGE POTENTIAL}

The construction of the $\pi \pi EP$ begins with the evaluation of 
the amplitude for on-shell NN scattering due to the exchange of two
pions. In order to avoid double counting, we must subtract the
term corresponding to the iterated OPEP and, on the centre of mass
of the NN system, the resulting amplitude is already the desired
potential in momentum space. As it depends strongly on the momentum 
transferred $\bbox{\Delta}$ and little on the nucleon energy E, we denote it
by  $T (\bbox{\Delta})$\footnote{The final relativistic expression for the 
amplitude depends also on powers of $\vec{z}=\vec{p}\,'+\vec{p}$,  which 
yield  ``non-local'' terms. We expand the amplitude in powers of 
$\vec{z}/m$ and keep just the first term, which gives the spin-orbit force.}.

In this work we are interested in the central, 
spin-spin, spin-orbit and tensor 
components of the configuration space potential,
which may be written in terms of local profile functions~\cite{Part70}.
They are related with the appropriate amplitudes in momentum space by

\begin{equation}
V(r) = - {\left(\frac{\mu}{2m}\right)}^2 {\frac{\mu}{4\pi}}
\int \frac{d^3 \bbox{\Delta}}{(2\pi)^3}\;
e^{-i\protect\bbox{\Delta} \protect\cdot \bbox{ r}}
\left[\left(\frac{4\pi }{\mu^3}\right) T (\bbox{\Delta})\right] .
\label{eq1} 
\end{equation}

In general, there are many processes that contribute to the $TPEP$.
However, for large distances, the potential is dominated by the low energy
amplitude for $\pi N$ scattering on each nucleon. When the external 
nucleons are on-shell,
the amplitude for the process $\pi^a(k) N(p) \rightarrow \pi^b(k')N(p')$
is written as

\begin{equation}
F = F^+ \, \delta_{ab} + F^- \, i \, \epsilon_{bac} \, \tau_c \ ,
\label{eq2}
\end{equation}

where
\begin{equation}
F^\pm = \overline u \left( A^\pm + \frac{{\not \! k} + {\not \! k}'}{2} \,
          B^\pm \right) u \ .
\label{eq3}
\end{equation}

The functions $A^\pm$ and $B^\pm$ depend on the variables 
\begin{equation}
t = \left(p-p'\right)^2\;\;\text{  and  }\;\;\nu = 
\frac{(p+p')\cdot (k+k')}{4m}
\end{equation}
\noindent or, alternatively, on 
\begin{equation}
s=(p+k)^2\;\;\text{  and  }\;\; u=(p-k')^2\;.
\end{equation}
\noindent 
When the pions are off-shell, they may also depend on
$k^2$ and ${k'}^2$. However, as discussed in Ref.~\cite{Rob96}, off-shell 
pionic effects have short range and do not contribute to the asymptotic 
amplitudes. At low energies, $A^\pm$ and $B^\pm$
may be written as a sum of
chiral contributions from the pure pion-nucleon sector,
supplemented by a series in the variables
$\nu$ and $t$\cite{HohI}, as follows:

\begin{eqnarray}
&& A^+ = \frac{g^2}{m} +\sum a^+_{mn} \, \nu^{2m} \, t^n  \ \ ,
\label{eq4}\\ [0.3cm]
&& B^+ = - \,\frac{g^2}{s-m^2} + \frac{g^2}{u-m^2}
+ \sum b^+_{mn} \, \nu^{(2m+1)} \, t^n \ \ ,
\label{eq5}\\ [0.3cm]
&& A^- =  \sum a^-_{mn} \, \nu^{(2m+1)} \, t^n \ \ ,
 \label{eq6}\\ [0.3cm]
&& B^- = -\, \frac{g^2}{s-m^2} - \frac{g^2}{u-m^2}
+ \sum b^-_{mn} \, \nu^{2m} \, t^n\ \ .  
\label{eq7}
\end{eqnarray}

\begin{figure}
\epsfxsize=8.5cm
\epsfysize=8cm
\centerline{\epsffile{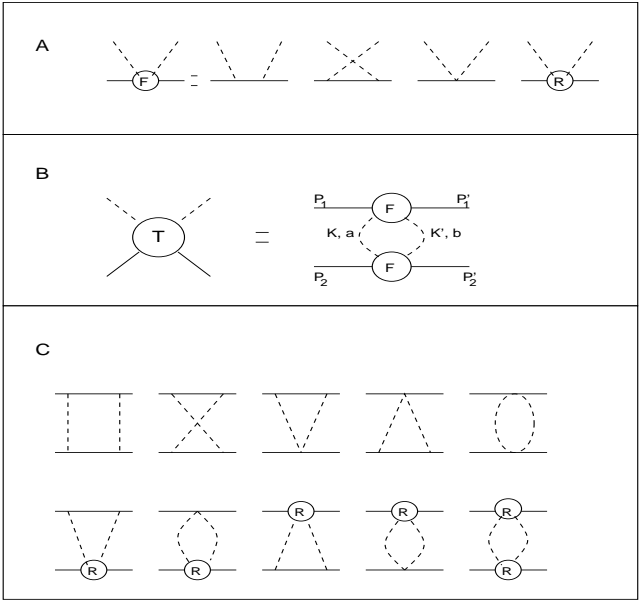}}
\vspace{0.4cm}
\caption{\it A) diagrams contributing to the low-energy $\pi N$ amplitude, where 
R represents the processes associated with the HJS coefficients; b) the 
two-pion exchange amplitude; c) contributions to the two-pion exchange 
amplitude from the purely pionic sector (top) and from processes involving 
the HJS coefficients (bottom).}
\label{Fig.1}
\end{figure}

\noindent
In these expressions, the nucleon contributions were calculated using a
non-linear pseudoscalar (PS) $\pi N$ coupling \cite{Rob96} and,  
in writing $A^+$, we have made explicit the factor $(g^2/m)$, 
associated with chiral symmetry. This amounts to just a redefinition
of  the usual $a_{00}^+$, given in Refs.~\cite{HohI,Hoh72}. 
On the other hand, the use of a 
pseudo vector (PV) $\pi N$ coupling would imply also a small redefinition of
$b_{00}^-$. It is very important to note, however, that the values of
the sub-amplitudes $A^\pm$ and $B^\pm$ are not at all influenced by 
this kind of choice and hence are completely model independent.
In the sequence, the terms in these expressions associated 
with the HJS coefficients will be denoted by
$A^\pm_R$ and $B^\pm_R$, the subscript $R$ standing for ``remainder'',
as indicated in Fig.~\ref{Fig.1}(A).


The evaluation of the diagrams of Fig.~\ref{Fig.1}(B) yields the
following general form for $T$

\begin{eqnarray}
T &=& -\frac{i}{2} \int \frac{d^4 Q}{(2\pi)^4} \, \frac{1}{k^2-\mu^2} \; 
\frac{1}{{k'}^2-\mu^2} \nonumber \\ [0.3cm] 
&\times&
 \left[ 3 F^{+(1)} \, 
   F^{+(2)} + 2 \bbox{\tau}^{(1)} \cdot \bbox{\tau}^{(2)} \, F^{-(1)} \, 
   F^{-(2)} \right]
\label{eq8}
\end{eqnarray}

\noindent
where the $F^{(i)}$ are given in Eq.~(\ref{eq2}) and the
factor $\case{1}/{2}$ accounts for the symmetry under the exchange of the
intermediate pions. The pion mass is represented by $\mu$ 
and the integration variable $Q$ is defined as

\begin{equation}
Q \equiv \frac{1}{2} \, (k+k') \ .                   \label{eq9}\\
\end{equation}

\noindent
In the sequence, we will also need the variables

\begin{eqnarray}
W &\equiv& p_1 + p_2 = p'_1 + p'_2 \ ,                 \label{eq10}\\ [0.3cm]
\Delta &\equiv& k' - k = p'_1 - p_1 = p_2 - p'_2 \ ,   \label{eq11}\\ [0.3cm]
z &\equiv& \frac{1}{2} \, \left[ (p_1 + p'_1) -
(p_2+p'_2) \right],                                  \label{eq12}\\ [0.3cm]
V_1 &\equiv& \frac{1}{2m} \; (W+z) \ ,                 \label{eq13}\\ [0.3cm] 
V_2 &\equiv& \frac{1}{2m} \; (W-z) .                   \label{eq14}
\end{eqnarray}

The evaluation of the diagrams of Fig.~\ref{Fig.1}(C) produces 

\begin{eqnarray}
&&T = -i\,(2m)^2 \, \frac{1}{2} \int \frac{d^4 Q}{(2\pi)^4} \; 
\nonumber \\ [0.3cm]
&\!\!\times&  \frac{1}{\left[ \left( Q -\frac{1}{2}\, \Delta \right)^2 - 
\mu^2 \right]  \left[ \left( Q + \frac{1}{2} \, \Delta \right)^2 - \mu^2
 \right]}  
  \nonumber \\ [0.3cm]
& \!\! \times & \!\! \left\{ 3 \left[ \left( \frac{g^2}{m} + A^+_R
  \right) I + \left( - \, \frac{g^2}{s-m^2} + \frac{g^2}{u-m^2} + 
   B^+_R  \right)
                \Qcor \right]^{(1)}\right. \nonumber \\ [0.3cm]
& \!\! \times &\left[ \left( \frac{g^2}{m} + A^+_R \right) I  
\left( - \,\frac{g^2}{s-m^2} + \frac{g^2}{u-m^2} 
  + B^+_R \right)  \Qcor \right]^{(2)} \nonumber \\ [0.3cm]
& \!\! +&  2\bbox{\tau}^{(1)} \cdot 
    \bbox{\tau}^{(2)} \left[
       A^-_R \,I + \left( - \,\frac{g^2}{s-m^2}-
\frac{g^2}{u-m^2} + B^-_R \right)
    \Qcor \right]^{(1)} \nonumber \\ [0.3cm]
& \!\! \times &\left. \left[ A^-_R \, I + \left( - \,
                 \frac{g^2}{s-m^2} - \frac{g^2}{u-m^2} + B^-_R \right) \Qcor
                  \right]^{(2)}  \right\},  
\label{eq15}
\end{eqnarray}

\noindent
where $I$ and $\Qcor$ are defined with non-relativistic
normalizations as

\begin{eqnarray}
&& I = \frac{1}{2m} \; \overline u u  \ ,  \label{eq16} \\[0.3cm]   
&& \Qcor = \frac{1}{2m} \; \overline u \, Q_\mu \, \gamma^\mu \, 
     u  \ . 
\label {eq17}
\end{eqnarray}

\noindent
The integrand also depends implicitly on $Q$
through the variables

\begin{eqnarray}
&& s_{i}- m^2 = Q^2 + Q \cdot (W\pm z) - \frac{1}{4} \; \Delta^2 \ ,
                                         \label{eq18} \\[0.2cm]
&& u_{i} - m^2 = Q^2 - Q \cdot (W\pm z) - \frac{1}{4} \; \Delta^2 \ ,
                                         \label{eq19}\\[0.2cm]
&& \nu_{i} = Q \cdot V_i \ .  \label{eq20}
\end{eqnarray}

\noindent
The integration is symmetric under the operation $\,Q \rightarrow - Q$
and hence nucleon denominators involving $s$ and $u$ yield identical
results.

The evaluation of the potential in configuration space requires also
an integration over $t$ and the pole structure
of Eq.~(\ref{eq8}) implies that the leading contribution
at very large distances comes from the region $t\approx 4\mu^2$
\cite{BrJack}, as it is well known.
 Therefore the form of our results in configuration
space becomes more transparent when the contribution of the HJS
coefficients is reorganized in terms of the dimensionless variable 

\begin{equation}
\theta \equiv \left(\frac{t}{4\mu^2}-1\right).
\label{eq21}
\end{equation}

\noindent
The amplitudes $A^\pm_R$ and $B^\pm_R$, associated with the
HJS coefficients, are rewritten as 

\onecolumn
\begin{eqnarray}
&& A^+_R = \frac{1}{\mu}
  \sum \alpha^+_{mn}\left(\frac{\nu}{\mu}\right)^{2m}\theta^n\; ,
  \label{eq23} \\ [0.2cm]
&& B^+_R = \frac{1}{\mu^2}
  \sum \beta^+_{mn}\left(\frac{\nu}{\mu}\right)^{(2m+1)}\theta^n \; ,
  \label{eq24} \\ [0.2cm]
&& A^-_R = \frac{1}{\mu}
  \sum \alpha^-_{mn}\left(\frac{\nu}{\mu}\right)^{(2m+1)}\theta^n \; ,
  \label{eq25} \\ [0.2cm]
&& B^-_R = \frac{1}{\mu^2}
  \sum \beta^-_{mn}\left(\frac{\nu}{\mu}\right)^{2m}\theta^n \; .
  \label{eq26} 
\end{eqnarray}

\noindent
In defining the coefficients $\alpha^\pm_{mn}$ and $\beta^\pm_{mn}$,
we have introduced powers of $\mu$ where appropriate so as to
make them dimensionless. Their numerical values are given in 
Tab.~\ref{Tab.1}.


\begin{table}
\caption{\it Values for the dimensionless coefficients of
Eqs.~(\protect\ref{eq23}-\protect\ref{eq26}) taken from 
Ref.~\protect\cite{HohI} and
re-stated by Eq.~\protect\ref{eq21}.}
\begin{tabular} {lcccccr}
{$(m,n)$}  & {$(0,0)$}  & {$(0,1)$}  & {$(0,2)$}  & {$(1,0)$}  &
{$(1,1)$}  & {$(2,0)$} \\ \tableline
$\alpha^+_{mn}$ & $  3.676\pm 0.138$ & $ 5.712\pm 0.096$ & $ 0.576\pm 0.048$ &
             $  4.62          $ & $-0.04          $ & $ 1.2  \pm 0.02 $ \\
$\beta^+_{mn}$ & $ -2.98 \pm 0.10 $ & $ 0.40 \pm 0.04 $ & $-0.16          $ &
             $ -0.68 \pm 0.06 $ & $ 0.32 \pm 0.04 $ & $-0.31 \pm 0.02 $ \\
$\alpha^-_{mn}$ & $-10.566\pm 0.212$ & $-1.976\pm 0.144$ & $-0.240\pm 0.032$ &
             $  1.222\pm 0.074$ & $ 0.208\pm 0.024$ & $-0.33 \pm 0.02 $ \\
$\beta^-_{mn}$ & $  9.730\pm 0.172$ & $ 1.760\pm 0.104$ & $ 0.40 \pm 0.032$ &
             $  0.86 \pm 0.07 $ & $ 0.22 \pm 0.02 $ & $ 0.25 \pm 0.02 $
\end{tabular}
\label{Tab.1}
\end{table}

Eq. (\ref{eq15}) can be naturally decomposed into a piece proportional 
to $g^4$, which originates in the pure pion-nucleon sector 
and a remainder, labelled by R, as in Fig.~\ref{Fig.1}(C). 
The former was discussed in
detail in Refs.~\cite{Rocha94,Rocha95}, where numerical expressions
were produced, and will no longer be considered here. We 
concentrate on $T_R$, which encompasses all the
other dynamical effects.

The potential in configuration
space may be written as

\begin{equation}
V_R = \left(V^+_{R1}+V^+_{R2}+V^+_{R3}+V^+_{R4}+V^+_{R5}
  +V^+_{R6}+V^+_{R7}+V^+_{R8}\right)
+\bbox{\tau}^{(1)} \cdot \bbox{\tau}^{(2)}
  \left(V^-_{R1}+V^-_{R2}+V^-_{R3}
  +V^-_{R4}+V^-_{R5}+V^-_{R6}+V^-_{R7}+V^-_{R8}\right)
\label{eq27}
\end{equation}

where the $V^\pm_{R_i}$ are integrals of the form

\begin{equation}
V^\pm_{Ri}  =  -\, {\left(\frac{\mu}{2m}\right)}^2 {\frac{\mu}{4\pi}}
\int \frac{d^3 \bbox{\Delta}}{(2\pi)^3}
e^{-i\protect\bbox{\Delta} \cdot \protect\bbox{ r}}
{\left[-i\frac{4\pi}{\mu^3}
\int \frac{d^4 Q}{(2\pi)^4} \;
\frac{1}{\left[ \left( Q -\frac{1}{2}\, \Delta \right)^2 - \mu^2 \right]
\left[ \left( Q + \frac{1}{2} \, \Delta \right)^2 - \mu^2 \right]}
\; g^\pm_i \right]},          \label{eq22}  
\end{equation}

\noindent
and the $g^\pm_i$ are the polynomials in $\nu/\mu$ and $\theta$ given in 
Appendix A. Thus we obtain the following general result for the $V^\pm_i$

\begin{eqnarray}
V^+_{R1}&=& -{\frac{\mu}{4\pi}}\frac{3}{2}\left\{g^2\frac{\mu}{m}
  \alpha^+_{mn} 2 S_{B(2m,n)} 
  +\alpha^+_{k\ell} \alpha^+_{mn}S_{B(2k+2m,\ell+n)}
  \right\}I^{(1)}I^{(2)},
  \label{eq28}\\ [0.3cm]
V^+_{R2}&=& -{\frac{\mu}{4\pi}}\frac{3}{2}
  \left\{g^2\frac{\mu}{m}\beta^+_{mn}S^{\mu}_{B(2m+1,n)}
  +\alpha^+_{k\ell}\beta^+_{mn}
  S^{\mu}_{B(2k+2m+1,n)}
  \right\} I^{(1)}\gamma_{\mu}^{(2)},
  \label{eq29}\\ [0.3cm]
V^+_{R4}&=& -{\frac{\mu}{4\pi}}\frac{3}{2}
  \left\{\beta^+_{k\ell} \beta^+_{mn}
   S^{\mu\nu}_{B(2k+2m+2,\ell+n)}
  \right\}\gamma_{\mu}^{(1)}\gamma_{\nu}^{(2)}, 
  \label{eq30}\\ [0.3cm]
V^-_{R1}&=& -{\frac{\mu}{4\pi}}
  \left\{\alpha^-_{k\ell} \alpha^-_{mn}S_{B(2k+2m+2,\ell+n)}
  \right\}I^{(1)}I^{(2)},
  \label{eq31}\\ [0.3cm]
V^-_{R2}&=& -{\frac{\mu}{4\pi}}
  \left\{\alpha^-_{k\ell} \beta^-_{mn}
   S^{\mu}_{B(2k+2m+1,\ell+n)}
  \right\}
  I^{(1)}\gamma_{\mu}^{(2)},
  \label{eq32}\\ [0.3cm]
V^-_{R4}&=& -{\frac{\mu}{4\pi}}
  \left\{\beta^-_{k\ell}\beta^-_{mn}
   S^{\mu\nu}_{B(2k+2m,\ell+n)}
  \right\}\gamma_{\mu}^{(1)}\gamma_{\nu}^{(2)},  
  \label{eq33}\\ [0.3cm]
V^+_{R5}&=& -{\frac{\mu}{4\pi}}\frac{3}{2}\frac{\mu}{m}
  \left\{g^2\alpha^+_{mn}S^{\mu}_{T(2m,n)}
   \right\}\gamma_{\mu}^{(1)}I^{(2)}, 
  \label{eq34}\\ [0.3cm]
V^+_{R7}&=& -{\frac{\mu}{4\pi}}\frac{3}{2}\frac{\mu}{m}
  \left\{g^2\beta^+_{mn}S^{\mu\nu}_{T(2m+1,n)}
  \right\}\gamma_{\mu}^{(1)}\gamma_{\nu}^{(2)},
  \label{eq35}\\ [0.3cm]
V^-_{R5}&=& {\frac{\mu}{4\pi}}\frac{\mu}{m}
  \left\{g^2\alpha^-_{mn}S^{\mu}_{T(2m+1,n)}
  \right\}\gamma_{\mu}^{(1)}I^{(2)},
  \label{eq36}\\ [0.3cm]
V^-_{R7}&=& {\frac{\mu}{4\pi}}\frac{\mu}{m}
  \left\{g^2\beta^-_{mn}S^{\mu\nu}_{T(2m,n)}
  \right\}\gamma_{\mu}^{(1)}\gamma_{\nu}^{(2)}.
  \label{eq37}
\end{eqnarray}

The expressions for $V^\pm_{R3}$, $V^\pm_{R6}$ and $V^\pm_{R8}$
are identical respectively to $V^\pm_{R2}$, 
$V^\pm_{R5}$ and $V^\pm_{R7}$ when
the very small differences between $\nu_1$ and $\nu_2$ are neglected. 
In these results $S_{B(m,n)}$ and $S_{T(m,n)}$ represent integrals 
of bubble (B) and triangle (T) diagrams, with $m$ and $n$ indicating the 
powers of $(\nu/\mu)$ and $\theta$ respectively, whose detailed form is
presented in appendix B. There, we show that the integrals with one 
free Lorentz index are proportional to $V^\mu_i$ whereas those with
two  indices may be proportional to either $V^\mu_iV^\nu_i$ or 
$g^{\mu\nu}$. Therefore we write for both bubble and triangle 
integrals

\begin{eqnarray}
S^\mu_{(m,n)}&=& V^\mu_iS^V_{(m,n)},
\label{eq38}\\ [0.3cm]
S^{\mu\nu}_{(m,n)}&=& V^\mu_iV^\nu_iS^{VV}_{(m,n)}+g^{\mu\nu}S^{g}_{(m,n)}.
\label{eq39}
\end{eqnarray}

Using the approximations described in Appendix B and the 
Dirac equation as in Eq.(B1), we obtain

\begin{eqnarray}
&&V^+_{R1}+V^+_{R5}+V^+_{R6}= -{\frac{\mu}{4\pi}}\frac{3}{2}
  \left\{g^2\frac{\mu}{m}
  \alpha^+_{mn} \left[2S_{B(2m,n)}+2S^V_{T(2m,n)}\right] 
+\alpha^+_{k\ell} \alpha^+_{mn}S_{B(2k+2m,\ell+n)}
  \right\}I^{(1)}I^{(2)},
  \label{eq40}\\ [0.4cm]
&&V^+_{R2}+V^+_{R3}+V^+_{R7}+V^+_{R8}= -{\frac{\mu}{4\pi}}\frac{3}{2}
  \left\{g^2\frac{\mu}{m}\beta^+_{mn}
  \left[2S^V_{B(2m+1,n)}+2S^{VV}_{T(2m+1,n)}\right] \right. \nonumber \\ [0.2cm]
&&+\left.\alpha^+_{k\ell}\beta^+_{mn}
  S^{V}_{B(2k+2m+1,n+\ell)}
  \right\} I^{(1)}I^{(2)} 
 -\ {\frac{\mu}{4\pi}}\frac{3}{2}
  \left\{g^2\frac{\mu}{m}\beta^+_{mn}
  2S^{g}_{T(2m+1,n)}\right\}
  \gamma^{(1)}\cdot\gamma^{(2)},
  \label{eq41}\\ [0.4cm]
&&V^+_{R4}= -{\frac{\mu}{4\pi}}\frac{3}{2}
  \left\{\beta^+_{k\ell} \beta^+_{mn}
   S^{VV}_{B(2k+2m+2,\ell+n)}\right\}I^{(1)}I^{(2)}
-{\frac{\mu}{4\pi}}\frac{3}{2}
  \left\{\beta^+_{k\ell} \beta^+_{mn}
  S^{g}_{B(2k+2m+2,\ell+n)}\right\}
  \gamma^{(1)}\cdot\gamma^{(2)}, 
  \label{eq42}\\ [0.4cm]
&&V^-_{R1}= -{\frac{\mu}{4\pi}}
  \left\{\alpha^-_{k\ell} \alpha^-_{mn}S_{B(2k+2m+2,\ell+n)}
  \right\}I^{(1)}I^{(2)},
  \label{eq43}\\ [0.4cm]
&&V^-_{R2}+V^-_{R3}= -{\frac{\mu}{4\pi}}
  \left\{\alpha^-_{k\ell} \beta^-_{mn}
   2S^{V}_{B(2k+2m+1,\ell+n)}
  \right\}
  I^{(1)}I^{(2)},
  \label{eq44}\\ [0.4cm]
&&V^-_{R4}= -{\frac{\mu}{4\pi}}
  \left\{\beta^-_{k\ell}\beta^-_{mn}
   S^{VV}_{B(2k+2m,\ell+n)}
  \right\}I^{(1)}I^{(2)}
-{\frac{\mu}{4\pi}}
  \left\{\beta^-_{k\ell}\beta^-_{mn}
   S^{g}_{B(2k+2m,\ell+n)}
  \right\}\gamma^{(1)}\cdot\gamma^{(2)}\, ,  
  \label{eq45}\\ [0.4cm]
&&V^-_{R5}+V^-_{R6}={\frac{\mu}{4\pi}}\frac{\mu}{m}
  \left\{g^2\alpha^-_{mn}2S^{V}_{T(2m+1,n)}
  \right\}I^{(1)}I^{(2)}\, ,
  \label{eq46}\\ [0.4cm]
&&V^-_{R7}+V^-_{R8}= {\frac{\mu}{4\pi}}\frac{\mu}{m}
  \left\{g^2\beta^-_{mn}S^{VV}_{T(2m,n)}
  \right\}I^{(1)}I^{(2)}
+{\frac{\mu}{4\pi}}\frac{\mu}{m}
  \left\{g^2\beta^-_{mn}S^{g}_{T(2m,n)}
  \right\}\gamma^{(1)}\cdot\gamma^{(2)}\, .  
  \label{eq47}
\end{eqnarray}

In configuration space, the spin-dependence
of the potential is obtained by means of the
non-relativistic results~\cite{Part70}
\begin{eqnarray}
 I^{(1)} \, I^{(2)} \; & \cong & \; 1 - \frac{\Omega_{S0}}{2m^2} \ \ ,
\label{eq48}\\ [0.3cm]
 \gamma^{(1)} \cdot \gamma^{(2)} & \cong & 
  1+3\frac{\Omega_{S0}}{2m^2}-\frac{\Omega_{SS}}{6m^2} 
  -\frac{\Omega_T}{12m^2} \ \ ,
\label{eq49}
\end{eqnarray}

\noindent
where
\twocolumn

\begin{eqnarray}
\Omega_{S0} & = &\bbox{ L \cdot S}\left( \frac{1}{r}\; 
  \frac{\partial}{\partial r} \right)  \ ,
\label{eq50}\\ [0.3cm]
\Omega_{SS} & = &- \bbox{\sigma}^{(1)} \cdot \bbox{\sigma}^{(2)} \left(
  \frac{\partial^2}{\partial r^2} + \frac{2}{r}\;
  \frac{\partial}{\partial r} \right) \ ,  
\label{eq51}\\ [0.3cm]
\Omega_T &\!\! = &\!\! \hat{S}_{12}  \left( \frac{\partial^2}{\partial r^2} \, -
  \frac{1}{r} \; \frac{\partial}{\partial r} \right) \ ,
\label{eq52}
\end{eqnarray}
\noindent and 
\[
\hat{S}_{12}=\left(3 \bbox{\sigma}^{(1)} \cdot \hat{\bf r} \,
  \bbox{\sigma}^{(2)} \cdot \hat{\bf r} - \bbox{\sigma}^{(1)} \cdot
  \bbox{\sigma}^{(2)} \right)\ .
\]

An interesting feature
of the partial contributions to the potential is that they are given
by two sets of phenomenological parameters, the $\pi N$ coupling 
constant and the HJS coefficients, multiplying structure integrals.
These integrals depend on just the pion and nucleon propagators and
hence carry very little model dependence. Their main features are 
discussed in the next section.


\section{INTEGRALS AND CHIRAL SYMMETRY}

Our expressions for the $TPEP$, given by Eqs.~(\ref{eq40}-\ref{eq47}),
contain both bubble and triangle integrals, which depend on the indices 
$m$ and $n$, associated respectively with the powers of $(\nu/\mu)$ and 
$\theta$, in the HJS expansion. The numerical evaluation of these integrals 
has shown that there is a marked hierarchy in their spatial behavior and that
the functions with $m=n=0$ prevail at large distances. 

\begin{figure}
\epsfxsize=7.0cm
\epsfysize=9.5cm
\centerline{\epsffile{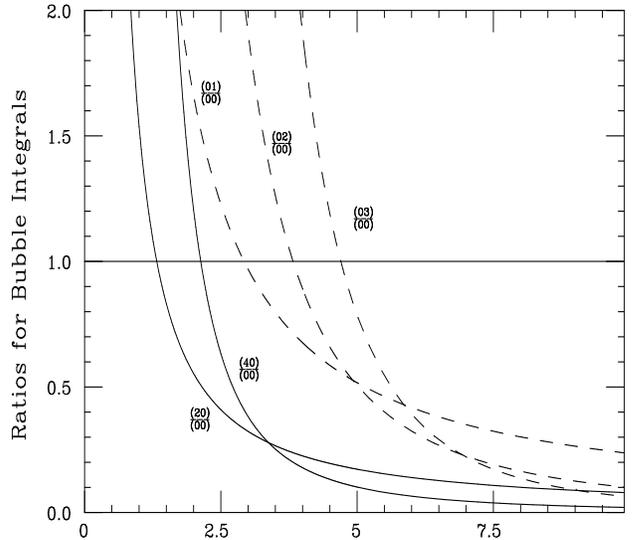}}
\caption{\it Asymptotic behavior of the bubble integrals $S_{B(m,n)}$.
The  ratios $S_{B(m,0)}/S_{B(0,0)}$ and $S_{B(0,n)}/S_{B(0,0)}$, for some
values of $m$ and $n$are indicated by solid and dashed lines respectively.
One sees that the integral 
$S_{B(0,0)}$ (unity line) is asymptotically dominant.}
\label{Fig.2}
\end{figure}

\begin{figure}
\epsfxsize=7.0cm
\epsfysize=9.5cm
\centerline{\epsffile{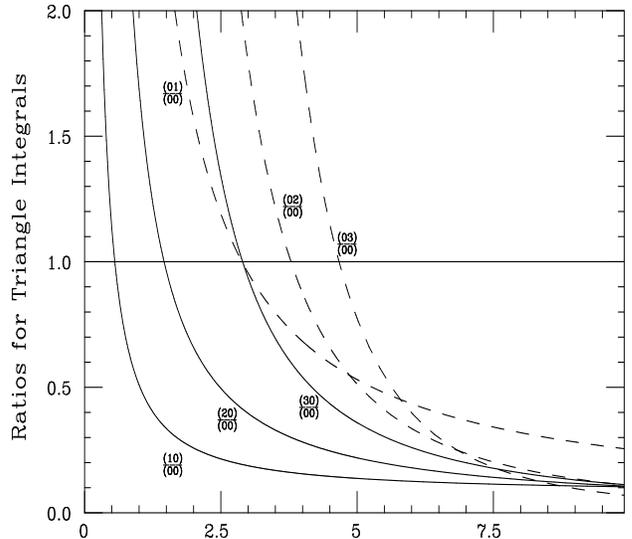}}
\caption{\it Asymptotic behavior of the triangle integrals $S_{T(m,n)}$. 
The  ratios $S_{T(m,0)}/S_{T(0,0)}$ and $S_{T(0,n)}/S_{T(0,0)}$, for some
values of $m$ and $n$are indicated by solid and dashed lines respectively.
As in 
Fig.~\protect\ref{Fig.2}, the integral for $m=n=0$ is asymptotically dominant.}
\label{Fig.3}
\end{figure}

In order to provide a feeling for the distance scales of the various effects, 
in Figs.~\ref{Fig.2} and~\ref{Fig.3} we display the ratios 
$\left[S_{B(m,n)}/S_{B(0,0)}\right]$ and 
$\left[S_{T(m,n)}^V/S_{T(0,0)}^V\right]$,  
for some values of $m$ and $n$, as functions of $r$.


When considering these figures, it is useful to bear in mind that the $(m,0)$
and $(0,n)$ curves convey different informations. The former series 
represents the average values of $\left(\nu /\mu\right)^m$ and is related 
to the behavior of the intermediate $\pi N$ amplitude below threshold. 
For physical
$\pi N$ scattering, the variable $\nu$ is always greater than $\mu$, whereas 
in the present problem the average values of $(\nu/\mu)^m$ are smaller than 1
for distances beyond 2.5 fm and tend to zero for very large values of $r$.
This is the reason why the construction of the 
$TPEP$ cannot be based on raw
scattering data, but rather, requires the use of dispersion relations in order
to transform the $\pi N$ amplitude to the suitable kinematical region
\cite{BrJack}. One has, therefore, a situation similar to the case of 
three-body forces, as discussed by Murphy and Coon~\cite{Murp95}, which 
emphasizes the role of the HJS coefficients.

Regarding the dependence of the integrals on the momentum transferred, one 
notes that the intermediate $\pi N$ amplitude in the momentum space is 
already in the physical $t<0$ region and does not require any extrapolations.
On the other hand, when one goes to configuration space, the Fourier 
transform picks up values of the amplitude around the point $t=4\mu^2$. 
Thus, the r-space potential is not transparent as far as $t$ is concerned and
the coherent physical picture only emerges when one uses it in the 
Schrodinger equation. This is a well known property, which also applies to
the OPEP. 

The fact that the integrals with  $m=n=0$ dominate at large distances
means that the
main contribution to the isospin symmetric central potential comes from 
Eq.~(\ref{eq40}) and is given by

\begin{eqnarray}
&&V^+_{R1}+V^+_{R5}+V^+_{R6}= -\frac{\mu}{4\pi}\,\frac{3}{2}
  \left\{g^2\,\frac{\mu}{m}\,\alpha^+_{00}\;\right. \nonumber \\ [0.3cm]
&& \times \left. 2\left[S_{B(0,0)}+S^V_{T(0,0)}\right]
+\left(\alpha^+_{00}\right)^2 S_{B(0,0)}
  \right\}I^{(1)}I^{(2)}.
  \label{eq53}
\end{eqnarray}

\begin{figure}
\epsfxsize=7.0cm
\epsfysize=7.0cm
\centerline{\epsffile{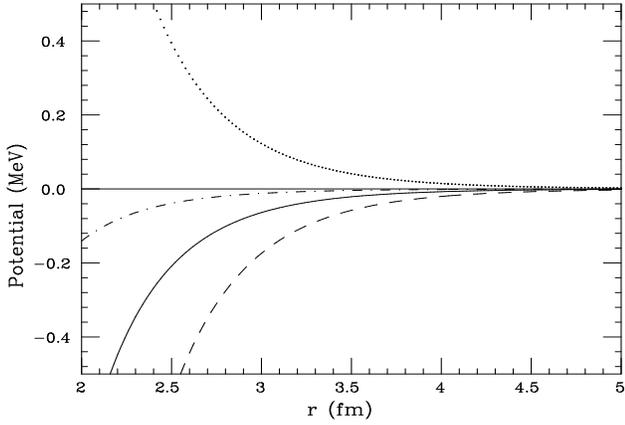}}
\caption{\it Structure of the leading contribution to the central potential, 
as given by Eq.~\protect\ref{eq53}. The continuous line represents the total 
effect, whereas the dashed, dotted, and dash-dotted lines correspond to the 
contributions proportional to $(g^2\mu/m)\alpha_{(00)}^+\,2S_B$, 
$(g^2\mu/m)\alpha_{(00)}^+\,2S_T$, and $\left(\alpha_{(00)}^+\right)^2\,S_B$
respectively.}
\label{Fig.4}
\end{figure}

The first term within curly brackets, proportional to $g^2$, is produced 
by the triangle and bubble diagrams in Fig.~\ref{Fig.1}(C)-bottom,
containing nucleons 
on one side and HJS amplitudes on the other, whereas the second one is due 
to the last diagram of Fig.~\ref{Fig.1}(C)-bottom. Inspecting Tab.~\ref{Tab.1} 
one learns that 
$\left(g^2\mu/m\right)/\alpha_{00}^+ \approx 8$, 
which suggests 
the first class of diagrams should dominate. On the other hand, the first
term is proportional to $\left[S_{B(0,0)}+S_{T(0,0)}^V\right]$ 
and, as discussed in
appendix B, these two integrals have opposite signs and there is a
partial cancellation between them. 
These features of the leading contribution are displayed in 
Fig.~\ref{Fig.4}, which shows that the first term is indeed dominant.


The cancellation noticed in the leading contributions is not a coincidence.
Instead, it represents a deep feature of the problem, which is due to chiral 
symmetry and also occurs in various other terms of the potential.

In appendix C we have shown that
the asymptotic form of $S_{B(0,0)}$ is given by the analytic expression

\begin{equation}
S_{B(0,0)}^{\text{\scriptsize asymp}} = \frac{1}{(4\pi)^2}\;
  2 \sqrt{\pi}\; \frac{e^{-2x}}{x^{5/2}}
  \left(1+\frac{3}{16}\frac{1}{x}
  -\frac{15}{512}\frac{1}{x^2}+\cdots\right)\; .
\label{eq54}
\end{equation}

\noindent
Its accuracy is 1\% up to 1.2 fm.
There, we also studied the form of the basic triangle integral $S_{T(0,0)}$
and have demonstrated that 
$S_{T(0,0)}^{\,\text{\scriptsize asymp}}= 
- S_{B(0,0)}^{\,\text{\scriptsize asymp}}$
when $(\mu/m)\rightarrow 0$. As the integrals with other values of $m$ and $n$
can be obtained from the leading ones, the same relationship holds for them as
well. This explains why Figs.~\ref{Fig.2} and~\ref{Fig.3} are so similar.

As we have discussed elsewhere~\cite{Ballot96a,Ballot96b}, 
important cancellations due to chiral symmetry also occur in the pure 
$\pi N$ sector. In order to stress this point, we have evaluated the 
contributions  of the diagrams in the top line of Fig.~\ref{Fig.1}(C), denoted
respectively by box $(\Box)$, crossed $(\Join)$, triangle $(\triangle)$ (twice)
and bubble $((\!))$, for three different values of the ratio $\mu/m$, namely
$$
\left(\frac{\mu}{m}\right)^{\text{exp}},\;\; \;
\frac{1}{10}\left(\frac{\mu}{m}\right)^{\text{exp}}, \;\;\text{ and }\;\;
\frac{1}{100}\left(\frac{\mu}{m}\right)^{\text{exp}}.
$$
\begin{figure}
\epsfxsize=7.0cm
\epsfysize=8.0cm
\centerline{\epsffile{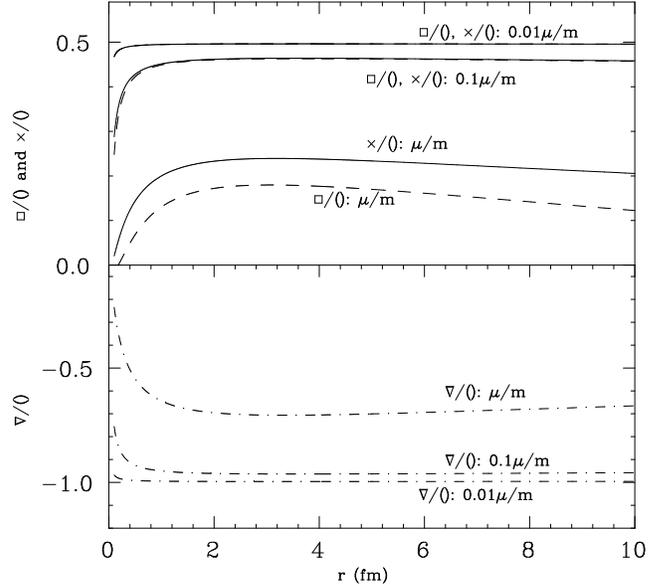}}
\vspace{0.5cm}
\caption{\it Contributions of the box, crossed, and triangle diagrams divided by 
that of the bubble, in the pure $\pi N$ sector, for the ratios of the pion over
the nucleon mass equal to the experimental value $\mu/m$, to $0.1\mu/m$, and 
to $0.01\mu/m$.}
\label{Fig.5}
\end{figure}
In Fig.~\ref{Fig.5} we display the ratios of the box, crossed and triangle
contributions over the bubble result as functions of distance, where it is 
possible to notice two interesting features. The first is that these 
ratios tend to become flat as $\mu/m$ decreases. The other one is that as 
$(\mu/m)\rightarrow 0$, one obtains the following relations:
$\Box=0.5 (\!)$, $\Join=0.5 (\!)$, and $\triangle=-(\!)$. Thus, for the 
amplitude in the pure $\pi N$ sector, we have $
\Box+\Join+2\triangle+(\!)=0$, a point also remarked by Friar and 
Coon~\cite{Fria94}. This result, when combined with the previous discussion 
concerning the bottom part of Fig.~\ref{Fig.1}(C), indicates that the 
two-pion exchange NN potential would vanish if chiral symmetry were exact, 
because the same would happen with the intermediate $\pi N$ amplitude. So,
all the physics associated with the tail of the intermediate range interaction
is due to chiral symmetry breaking.

As a final comment, we would like to point out that in the evaluation of 
the $TPEP$ there are two different hierarchies that can be used to
simplify calculations. One of them concerns the HJS coefficients, which
are more important for low powers on $\nu$ and $t$. The other one is associated 
with the spatial behavior of the integrals as functions of $m$ and $n$. 
The combined use of these hierarchies allow many terms to be discarded.


\section{RELATED WORKS}

To our knowledge, only Ord\'o\~nez, Ray and van Kolck have so far attempted 
to derive realistic nucleon-nucleon phenomenology in the framework of 
chiral symmetry~\cite{Ordo94,Ordo96}. 
The potential obtained by these
authors is based on a very general
effective Lagrangian, which is approximately invariant
under chiral symmetry to a 
given order in non-relativistic momenta and pion mass. They considered
explicitly the degrees of freedom associated with pions, nucleons and
deltas, whereas the effects of other interactions were incorporated into
parameters arising from contact terms and higher order derivatives.
In principle the free parameters in their effective Lagrangian
could be obtained from other physical processes, but at present
only some of them are known\footnote{See Ref.~\cite{Ber95} for a comprehensive 
discussion of this point.}. In their work these parameters were obtained 
by fitting deuteron properties and NN
observables for $j\le 2$ whereas loop integrals
were regularized by means of non-covariant Gaussian cutoffs
of the order of the $\rho$ meson mass. 
Thus they could show that the effective chiral 
Lagrangian approach is flexible enough for allowing 
the data to be reproduced with an appropriate choice of 
dynamical parameters and cutoffs.
Comparing their approach to ours, one notes several important
differences. For instance, we use dimensional regularization which is
well known to preserve the symmetries of the problem and our expressions 
are quite insensitive to short distance effects. In the work of Ord\'o\~nez, 
Ray and van Kolck, on the other hand, 
``variations in the cutoff are compensated to some extent by a redefinition 
of the free parameters in the theory.''~\cite{Ordo96}.
Moreover, we use the HJS coefficients as input, which are determined by 
$\pi N$ scattering, and therefore our results yield predictions for 
interactions at large distances or, alternatively,
for $j\ge 2$. The test of these predictions will be presented elsewhere.

Another point in the present work that deserves to be discussed concerns 
the subtraction of the iterated OPEP. In our calculation of the  
$TPEP$ in the pure nucleonic sector, we
have supplemented the results derived by Lomon and Partovi 
~\cite{Part70} for the pseudoscalar
box and crossed box diagrams with bubble and triangle 
diagrams associated with chiral symmetry\cite{Rocha94}. 
We have also shown that the
use of a pseudovector coupling yields exactly the same results
and hence that the potential does not depend on how the symmetry
is implemented. However, the Partovi and Lomon 
amplitude include
the subtraction of the OPEP by means of the Blankenbecler-Sugar 
reduction of the relativistic equation and hence our results are
also affected by that procedure. This kind of 
choice should not influence measured quantities, since it 
amounts to just a selection of the conceptual basis to treat the problem
~\cite{Desp92}.
As discussed by Friar ~\cite{Fri77}
and more recently by Friar and Coon~\cite{Fria94}, the treatments of
the iterated OPEP by
Taketani, Machida and Ohnuma\cite{TMO}
and by Brueckner and Watson\cite{BW} differ by terms which are 
energy dependent. However, in our calculation,
energy dependent terms can be translated into the variable $\nu$ and,
in the previous section, we have
shown that the $TPEP$ at large distances
is dominated by the region where $\nu \approx 0$. Hence
our results are not affected by the way
the OPEP is defined. Another indication that confirms this fact
comes from two recent studies dealing with the  relative weights
of the various $TPEP$ contributions to NN phase shifts, 
which have shown that the
role of the iterated OPEP is very small for $j\ge 2$
\cite{Ballot96a,Ballot96b}.

The last comment we would like to make in this section concerns the 
dynamical significance of the HJS coefficients.
It has long been known
that a tree model for the intermediate $\pi N$ amplitude containing
nucleons, deltas, rho mesons and an amplitude describing the 
$\sigma$-term can be made consistent with the
experimental values of the HJS coefficients by means of a rather 
conservative choice of masses and coupling constants
~\cite{HohI,Murp95,Olson75,Sca74,Mene85}.
In general, there are two advantages of employing such a model
in a nuclear physics calculation. The first is that it allows one to go
beyond the HJS coefficients, specially as far as the the pion off-shell
behaviour of the amplitude is concerned. However, as we have discussed above,
this kind of off-shell effects are related to short distance interactions
and hence are not important for the asymptotic $TPEP$.
It is in this sense that we consider our results to be
model independent.
The second motivation for using a model 
is that it may provide a dynamical picture involving the various 
degrees of freedom of the problem and shed light into their relative 
importance.
As we show in the next section, the leading contribution to the
scalar-isoscalar potential comes from the coefficient
$\alpha^+_{00}\equiv\mu(a^+_{00}+4\mu^2a^+_{01}+16\mu^4a^+_{02})$.
As expected, it is attractive and determined mostly by the
$\,\pi N\,$ sigma term and by the delta. The former 
yields $\alpha^+_{00\Sigma}= 1.8$ whereas the latter is
the outcome of a strong cancellation between pole and non-pole contributions
$\alpha^+_{00\Delta}=(26.5-25.2)$\cite{HohI}. Thus the delta non-pole 
term plays a very important role in the interaction, and 
must be carefully considered in any model aiming at being realistic. 


\section{RESULTS AND CONCLUSIONS}

In this work we have assumed that the $TPEP$ is due to both pure 
pion-nucleon interactions and processes involving other degrees of freedom,
as represented in the top and bottom lines of Fig.~\ref{Fig.1}(C).
The former class of processes was evaluated and studied 
elsewhere~\cite{Rocha94,Ballot96b} and hence we here concentrate on the latter.


\begin{figure}
\epsfxsize=6.3cm
\epsfysize=6.0cm
\centerline{\epsffile{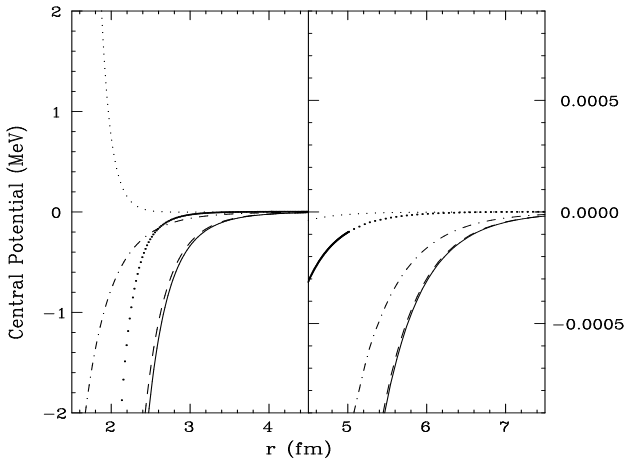}}
\vspace{0.5cm}
\caption{\it Structure of the central potential; the dot-dashed curve represents 
the leading contribution (Eq.~(\protect\ref{eq53})) whereas the dashed,
big dotted and small dotted curves correspond to 
Eqs.~(\protect\ref{eq40}), (\protect\ref{eq41}), and (\protect\ref{eq42})
respectively; the solid line represent the full potential.}
\label{Fig.6}
\end{figure}

As discussed in Sec. III, the leading contribution to the potential at large 
distances is due to the intermediate $\pi N$ amplitude around 
the point $\nu=0$, $t=4\mu^2$. In order to understand the role played
by the other terms, in Fig.~\ref{Fig.6} we disclose the structure of the 
scalar-isoscalar potential, given by Eqs.~(\ref{eq40}-\ref{eq42}). There
it is possible to see that Eq.~(\ref{eq40}), associated with the 
$\alpha^+_{m n}$ HJS 
coefficients, completely dominates the full potential. On the other hand,
for moderate distances, there is a clear separation between the curves 
representing the leading contribution, given by Eq.~(\ref{eq53}), and the total
potential. This indicates that corrections associated with higher powers of 
$\nu$ and $t$ are important there, a feature that could have been anticipated 
from Figs.~\ref{Fig.2} and~\ref{Fig.3}.


\begin{figure}
\epsfxsize=6.3cm
\epsfysize=5.8cm
\centerline{\epsffile{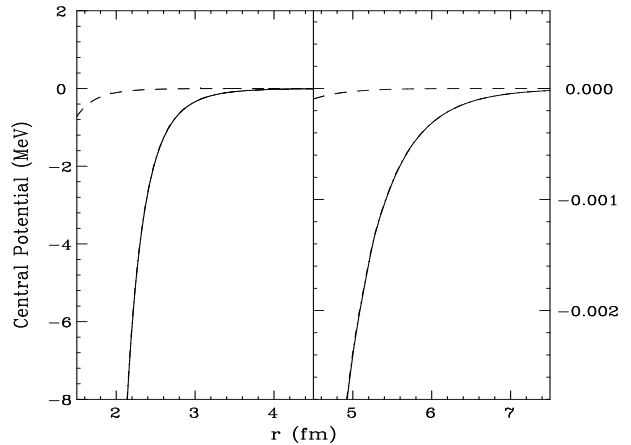}}
\vspace{0.3cm}
\caption{\it Contributions for the total $TPEP$, represented by continuous 
line; the dashed line comes from the pure $\pi N$ sector 
(Fig.~\protect\ref{Fig.1}(C)-top),
whereas that associated with other degrees of freedom falls on top of the 
continuous line and cannot be distinguished from it.}
\label{Fig.7}
\end{figure}

The total potential, obtained by adding the results of 
Refs.~\cite{Rocha94,Rocha95} with those of this work, is given in 
Fig.~\ref{Fig.7}, where it is possible to see that the contribution from 
the pure nucleon sector is rather small. This information, when combined with 
those contained in the preceding figures, allows one to conclude that the 
strength of the scalar-isoscalar attraction at large distances is due mostly 
to diagrams involving the nucleon on one side and the remaining 
degrees of freedom on the other.


\begin{figure}
\epsfxsize=6.3cm
\epsfysize=5.8cm
\centerline{\epsffile{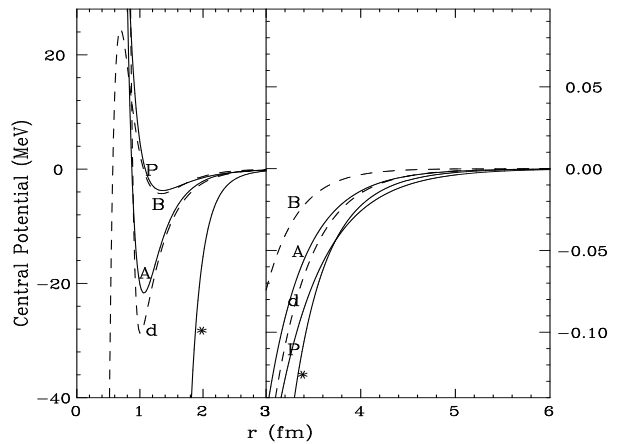}}
\vspace{0.3cm}
\caption{\it Central components of various potentials: 
parametrized Paris~\protect\cite{Laco80} (solid, P),
Argonne v14~\protect\cite{WSA84} (solid, A),
dTRS~\protect\cite{TRS75} (dashed, d), 
Bonn~\protect\cite{MHE87} (dashed, B), 
and our full potential (solid, *).}
\label{Fig.8}
\end{figure}

In Fig.~\ref{Fig.8} we compare our results for the scalar-isoscalar
interaction with the corresponding  
components of some potentials found in the literature: 
parametrized Paris~\cite{Laco80}, 
Argonne v14~\cite{WSA84}, dTRS~\cite{TRS75}, and Bonn~\cite{MHE87}. 
The first thing that 
should be noted is that all curves but ours bend upwards close to the 
origin, indicating clearly that the validity of our results is restricted 
to large distances. Inspecting the medium and long distances regions, it is 
possible to see that every potential disagrees with all the others. 
On the other 
hand, this does not prevent the realistic potentials from reproducing 
experimental data, something that is possible because there is a compensation
arising from the other discrepancies found in the short distance region.
It is for this reason that the 

\onecolumn
\noindent
accurate knowledge of the tail of the potential
may yield indirect constraints over its short distance part.

\begin{figure}
\epsfxsize=12.0cm
\epsfysize=10.0cm
\centerline{\epsffile{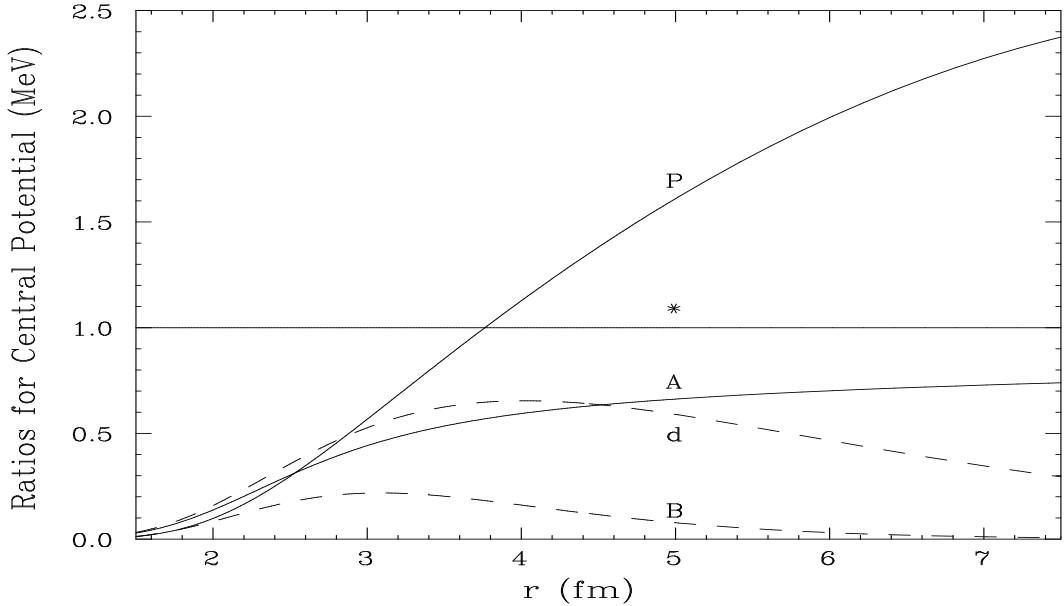}}
\caption{\it Ratio of the central components of some realistic potentials 
by our full result (solid,*): 
parametrized Paris~\protect\cite{Laco80} (solid, P),
Argonne v14~\protect\cite{WSA84} (solid, A),
dTRS~\protect\cite{TRS75} (dashed, d), and
Bonn~\protect\cite{MHE87} (dashed, B).}
\label{Fig.9}
\end{figure}

Finally, in Fig.~\ref{Fig.9} we show the ratios of the realistic potentials 
by our full potential, where the discrepancies mentioned above appear again,
in a different form. An interesting feature of this figure is that the 
realistic potentials come close together around 2 fm, suggesting that this 
region is important for reproduction of experimental data. Moreover, all of 
them show inflections there, indicating that the physics in this region goes 
beyond the exchange of two 
uncorrelated pions. In the long distance domain, the $r$ dependence of the
Argonne potential is not too different from ours, because it is based on a 
square OPEP form.


In summary, in this work we have shown that the use of a chiral $\pi N$
amplitude, supplemented by experimental information, determines uniquely 
the long-distance features of the scalar-isoscalar component of the $NN$ 
potential. As it is well known, the kinematical regions relevant to this 
problem  are not directly accessible by experiment and hence empirical 
information has to be treated by means of dispersion relations before 
being used as input in the calculations of the force. 
From a purely mathematical point of view, our results are valid for 
$r >2.5$ fm, since in this region one has $\nu <\mu$ and the HJS coefficients
may be safely employed. On the other hand, the determination of the dynamical 
validity of the results is much more difficult, since this requires a 
comparision with processes involving the mutual interaction of the 
exchanged pions, something that remains to be done in the framework of chiral 
symmetry. 

In general, a potential involves two complementary ingredients that deserve 
attention, namely geometry and dynamics. In our calculation, 
the former is associated with 
standard bubble and triangle integrals, that determine unambiguously the 
profile functions in configuration space, whereas 
dynamics is incorporated into the problem by means of 
coupling constants and empirical 
coefficients. Geometry and dynamics decouple in our
final expressions and hence they would remain valid even if  changes in the 
values of the dynamical constants may occur 
in the future. In the case of Fig.~\ref{Fig.9}, such a 
change would amount to just a modification of the vertical scale, with no 
appreciable effect on the discrepancies found with phenomenological 
potentials.

\section{ACKNOWLEDGEMENTS}
M.R.R. would like to thank the kind hospitality of Nuclear Theory Group of 
the Department of Physics of the University of Washington, Seattle, during
the performance of this work. This work was partially supported 
by U.S. Department of Energy. The work of C.A. da Rocha was supported by
CNPq, Brazilian Agency.


\appendix
\section{Functions \ $\lowercase{g}^\pm_{\imath}$}

We present here the polynomials that enter Eq.~(\ref{eq22}).
The groups of indices $i=1...4$ and $5...8$ refer, respectively
to bubble and triangle diagrams.


\begin{eqnarray}
g^+_1&=& 6\frac{m^2}{\mu^2}\left\{g^2\frac{\mu}{m}
  \alpha^+_{mn}\left[\left(\frac{\nu_1}{\mu}\right)^{2m} 
  +\left(\frac{\nu_2}{\mu}\right)^{2m}\right]\theta^n 
+\alpha^+_{k\ell} \alpha^+_{mn}
  \left(\frac{\nu_1}{\mu}\right)^{2k}\left(\frac{\nu_2}{\mu}\right)^{2m}
  \theta^{(\ell+n)}\right\}I^{(1)}I^{(2)},
  \label{eqA1}\\ [0.4cm]
g^+_2&=& 6\frac{m^2}{\mu^2}
  \left\{g^2\frac{\mu}{m}\beta^+_{mn}
  \left(\frac{\nu_2}{\mu}\right)^{(2m+1)}\theta^n
+\alpha^+_{k\ell}\beta^+_{mn}\left(\frac{\nu_1}{\mu}\right)^{2k}
  \left(\frac{\nu_2}{\mu}\right)^{(2m+1)}\theta^{(\ell + n)}\right\}
  \frac{1}{\mu}I^{(1)}\Qcor^{(2)},
  \label{eqA2}\\[0.4cm]
g^+_4&=& 6\frac{m^2}{\mu^2}\left\{ \beta^+_{k\ell} \beta^+_{mn}
  \left(\frac{\nu_1}{\mu}\right)^{(2k+1)} \left(\frac{\nu_2}{\mu}\right)^{(2m+1)}
  \theta^{(\ell+n)} \right\}
  \frac{1}{\mu^2}\Qcor^{(1)}\Qcor^{(2)}, 
  \label{eqA3}\\ [0.4cm] 
g^-_1&=& 4\frac{m^2}{\mu^2}
  \left\{\alpha^-_{k\ell} \alpha^-_{mn}\left(\frac{\nu_1}{\mu}\right)^{(2k+1)}
  \left(\frac{\nu_2}{\mu}\right)^{(2m+1)}\theta^{(\ell+n)}\right\}  
  I^{(1)}I^{(2)},
  \label{eqA4}\\ [0.4cm] 
g^-_2&=& 4\frac{m^2}{\mu^2} \left\{\alpha^-_{k\ell}
  \beta^-_{mn}\left(\frac{\nu_1}{\mu}\right)^{(2k+1)}
  \left(\frac{\nu_2}{\mu}\right)^{2m}\theta^{(\ell+n)} \right\}
  \frac{1}{\mu}I^{(1)}\Qcor^{(2)},
  \label{eqA5}\\[0.4cm]
g^-_4&=& 4\frac{m^2}{\mu^2}\left\{\beta^-_{k\ell}\beta^-_{mn}
  \left(\frac{\nu_1}{\mu}\right)^{2k} \left(\frac{\nu_2}{\mu}\right)^{2m}
  \theta^{(\ell+m)} \right\}
  \frac{1}{\mu^2}\Qcor^{(1)}\Qcor^{(2)},  
  \label{eqA6}\\[0.4cm]
g^+_5&=& \frac{2m\mu}{Q^2-2m Q\cdot V_1-\frac{1}{4}\Delta^2}\; 
  6\frac{m}{\mu}\left\{g^2\alpha^+_{mn}\left(\frac{\nu_2}{\mu}\right)^{2m}\theta^n 
  \right\}\frac{1}{\mu}\Qcor^{(1)}I^{(2)}, 
  \label{eqA7}\\[0.4cm]
g^+_7&=& \frac{2m\mu}{Q^2-2m Q\cdot V_1-\frac{1}{4}\Delta^2}\; 
  6\frac{m}{\mu}\left\{g^2\beta^+_{mn}\left(\frac{\nu_2}{\mu}\right)^{(2m+1)}\theta^n 
  \right\}\frac{1}{\mu^2}\Qcor^{(1)}\Qcor^{(2)},
  \label{eqA8}\\[0.4cm]
g^-_5&=& \frac{-2m\mu}{Q^2-2m Q\cdot V_1-\frac{1}{4}\Delta^2}\; 
  4\frac{m}{\mu}\left\{g^2\alpha^-_{mn}\left(\frac{\nu_2}{\mu}\right)^{(2m+1)}\theta^n
  \right\}\frac{1}{\mu}\Qcor^{(1)}I^{(2)},
  \label{eqA9}\\[0.4cm]
g^-_7&=& \frac{-2m\mu}{Q^2-2m Q\cdot V_1-\frac{1}{4}\Delta^2}\; 
  4\frac{m}{\mu}\left\{g^2\beta^-_{mn}\left(\frac{\nu_2}{\mu}\right)^{2m}\theta^n
  \right\}\frac{1}{\mu^2}\Qcor^{(1)}\Qcor^{(2)}
  \label{eqA10}
\end{eqnarray}

The expressions for $g^\pm_3$, $g^\pm_6$ and $g^\pm_8$
are obtained respectively from $g^\pm_2$, $g^\pm_5$ and $g^\pm_7$
by exchanging $\nu_1$ and $\nu_2$.


\section{INTEGRALS}

In this appendix we present the expressions for the integrals 
$S_{B(m,n)}$ and $S_{T(m,n)}$ that determine the potential
given in Sec. II. In many cases, a considerable simplification 
of the results, with no loss of numerical accuracy, can
be achieved due to the fact that one is interested in the 
asymptotic behaviour of the potential in configuration space. 
This allows one to ignore contact terms associated
with delta-functions or, alternatively, 
constant terms in momentum space integrals.

In bubble integrals the denominators involve just two pion
propagators, whereas there is an extra nucleon propagator
for triangles. In both cases, the integrands have
the general form of a polynomial in the variables 
$\frac{\nu_i}{\mu} = \left(\frac{Q}{\mu}\cdot V_i\right)$, 
where $V_i = \frac{1}{2m}(W\pm z)$.
In elastic pion-nucleon scattering at low energies,
$\nu=\mu$ at threshold and hence $\mu$ is also a \hfill

\twocolumn
\noindent
natural unit
for the $\nu_i$. In this problem,
$W\approx 2m$, $z\approx p$ and hence $V_i \approx 1$.
Moreover, using the mass-shell condition for the external
nucleons, we obtain

\begin{eqnarray}
\not V^{(i)} &=& I^{(i)},\\ [0.2cm]
\label{eqB1}
V_i^2 &=& 1-\frac{\Delta^2}{4m^2},\\[0.2cm]
\label{e
qB2}
V_{1} \cdot V_{2} &=& 1-\frac{\Delta^2}{4m^2}-\frac{z^2}{2m^2}.
\label{eqB3}
\end{eqnarray}

\noindent
These results show that the differences between $\nu_1$ and 
$\nu_2$ are of the order of relativistic corrections and therefore 
may be neglected.

In our expression for the potential, Lorentz tensors
proportional to $\Delta's$ always appear contracted
to either $V_i$ or $\gamma$ matrices. The use of the
equations of motion imply in the vanishing of these products
and hence we do not write them explicitly below.
We also make use of the result $\,\Delta^2 = t = - \bbox{\Delta}^2$.

When going to configuration space, it is useful to use the following
representation for the logarithm

\begin{equation}
\ln \; \left[1+\frac{\bbox{\Delta}^2}{M^2}\right] =
 - \int_0^1 d\gamma \;\frac{M^2}{\gamma^2}\;
\frac{1}{\bbox{\Delta}^2+\frac{M^2}{\gamma}},
\label{eq.B4}
\end{equation}

\subsection{Bubble integrals in momentum space:}

The basic bubble integral is

\begin{equation}
I_B^{\mu \ldots \sigma} = \int \frac{d^4 Q}{(2\pi)^4}\;  
\frac{\frac{Q^\mu}{\mu} \ldots \frac{Q^\sigma}{\mu}}
{\left[(Q-\frac{\Delta}{2})^2 -\mu^2\right]\left[(Q+\frac{\Delta}{2})^2 -\mu^2
\right]}.  
\label{eqB5}
\end{equation}

\noindent
The symmetry of the integrand makes all integrals with odd powers
of $Q$ to vanish.

The simplest case corresponds to
\begin{equation}
I_B = \int \frac{d^4 Q}{(2\pi)^4}\;
\frac{1}{\left[(Q-\frac{\Delta}{2})^2 -\mu^2\right]\left[(Q+\frac{\Delta}{2})^2
 -\mu^2\right]}. 
\label{eqB6}
\end{equation}

\noindent
Using Feynman integration parameters we write

\begin{equation}
I_B = \int^1_0 d\alpha \int \frac{d^4 Q}{(2\pi)^4} \; 
\frac{1}{\left[ Q^2 + 2P \cdot Q - M^2 \right]^2},
\label{eqB7}
\end{equation}

where
\begin{eqnarray}
P &\!\! = & \!\!  \left( \alpha - \frac{1}{2} \right) \Delta  \ ,
   \label{eqB8} \\ [0.2cm]
M^2 &\!\! = & \!\! \mu^2 - \frac{1}{4} \, \Delta^2 \ .
   \label{eqB9}    
\end{eqnarray}

Using the technique of dimensional regularization described in 
\cite{Rob96}, the integration over Q yields, after dropping the 
constant and divergent terms

\begin{equation}
I_B = - \frac{i}{(4\pi)^2} \int^1_0 d\alpha 
\ln \left[1+\frac{\bbox{\Delta}^2}{M_B^2} \right], 
\label{eqB10}
\end{equation}

\noindent
where 

\begin{equation}
M_B^2 = \frac{\mu^2}{\alpha (1-\alpha)}.
\label{Eq.11}
\end{equation}

Using Eq.~(B4), we obtain

\begin{equation}
I_B = \frac{i}{(4\pi)^2}
\int^1_0 d\alpha \int_0^1 d\beta \;
\frac{M_B^2}{\beta^2}\;
\frac{1}{\bbox{\Delta}^2+\frac{M_B^2}{\beta}}.
\label{Eq.12}
\end{equation}

For the integral $I_B^{\mu\nu}$, the same procedure yields

\begin{equation}
I_B^{\mu\nu}= \frac{1}{2} \, g^{\mu\nu}
\frac{i}{(4\pi)^2} 
\int^1_0 d\alpha \int_0^1 d\beta \;
\left(1-\frac{1}{\beta}\right)\;
\frac{M_B^2}{\beta^2}\;
\frac{1}{\bbox{\Delta}^2+\frac{M_B^2}{\beta}},
\label{Eq.13}
\end{equation}

\noindent
neglecting terms proportional 
to $\Delta^{\mu} \Delta^{\nu}$.

\vspace{0.3cm}

Analogously, for $I_B^{\mu\nu\rho\lambda}$, we have

\begin{eqnarray}
I_B^{\mu\nu\rho\sigma}&=& \frac{1}{8} \,
\left(g^{\mu\nu}g^{\rho\sigma}+g^{\nu\rho}g^{\mu\sigma}
+g^{\mu\rho}g^{\nu\sigma}\right)
\frac{i}{(4\pi)^2} \nonumber \\ [0.3cm]
&\times&\int^1_0 d\alpha \int_0^1 d\beta \;
\left(1-\frac{1}{\beta}\right)^2\;
\frac{M_B^2}{\beta^2}\;
\frac{1}{\bbox{\Delta}^2+\frac{M_B^2}{\beta}}.
\label{Eq.14}
\end{eqnarray}

\subsection{Triangle integrals in momentum space:}

The triangle integrals have the structure

\begin{eqnarray}
I_T^{\mu \ldots \sigma}&=&\int \frac{d^4 Q}{(2\pi)^4}\;\frac{2m\mu\;\; 
\frac{Q^\mu}{\mu} \ldots \frac{Q^\sigma}{\mu}}
{\left[Q^2-2m V_i \cdot Q-\frac{1}{4}\Delta^2 \right]}  \nonumber \\ [0.3cm]
&\times&\frac{1}
{\left[(Q-\frac{\Delta}{2})^2 -\mu^2\right]\left[(Q+\frac{\Delta}{2})^2 -\mu^2
\right]} .
\label{eqB15}
\end{eqnarray}

The basic case is

\begin{eqnarray}
I_T^\mu& = &\int \frac{d^4 Q}{(2\pi)^4} \; \frac{2m\mu\;\;\frac{Q^\mu}{\mu}}
{\left[Q^2-2m V_i \cdot Q-\frac{1}{4}\Delta^2 \right]}\nonumber \\ [0.3cm]
&\times&\frac{1}{\left[(Q-\frac{\Delta}{2})^2 -\mu^2\right]\left[(Q+
\frac{\Delta}{2})^2 -\mu^2\right]},
\label{eqB16}
\end{eqnarray}

\noindent
which corresponds to

\begin{eqnarray}
I_T^\mu &=& 2 \int^1_0 d\alpha (1-\alpha)\;\int^{1}_0 d\beta \nonumber \\
[0.2cm]
&\times&  \int \frac{d^4 Q}{(2\pi)^4} \; 
  \frac{2m\mu\;\;\frac{Q^\mu}{\mu}}
  {\left[Q^2+2P\cdot Q-M^2\right]^3},
\label{eqB17}
\end{eqnarray}

\onecolumn
\noindent
with

\begin{eqnarray}
P & = &-\frac{1}{2}\left[\alpha-(1-\alpha)\beta\right]\Delta - 
(1-\alpha)(1-\beta)mV_i ,
\label{eqB18} \\
M^2 & = & \left[\alpha+(1-\alpha)\beta\right]\mu^2 
  +\left[1-2\alpha-2(1-\alpha)\beta\right]\frac{\Delta^2}{4}.
\label{eqB19}
\end{eqnarray}

Integrating over $Q$, we have

\begin{equation}
I^\mu_T = -V_i^{\mu}\frac{i}{(4\pi)^2} \;
\left(\frac{m}{\mu}\right)\int^1_0 d\alpha \;\frac{(1-\alpha)}{\alpha}\;
\int^{1}_{0} d\beta \; \frac{1-\beta}{\beta}\;
\frac{2m\mu}{\bbox{\Delta}^2 + M_T^2},
\label{eqB20}
\end{equation}

\noindent
where

\begin{equation}
M_T^2 = \frac{\left[\alpha+(1-\alpha)\beta\right]\mu^2
+\left[(1-\alpha)(1-\beta)\right]^2m^2}
{\alpha(1-\alpha)\beta}.
\label{eqB21}
\end{equation}

Using the same procedure, and neglecting divergent terms, we get

\begin{eqnarray}
I^{\mu\nu}_T  &=& -V^\mu_i V^\nu_i
  \frac{i}{(4\pi)^2}\left(\frac{m}{\mu}\right)^2
  \int^1_0 d\alpha \frac{(1-\alpha)^2}{\alpha}\;
  \int^{1}_0 d\beta \frac{(1-\beta)^2}{\beta}\;
  \frac{2m\mu}{\bbox{\Delta}^2 + M^2_T} 
  \nonumber \\ [0.4cm]
&+& g^{\mu\nu}\frac{i}{(4\pi)^2}\left(\frac{m}{\mu}\right)
  \int^1_0 d\alpha (1-\alpha)\;
  \int^{1}_0 d\beta 
  \int_0^1 d\gamma \;
  \frac{M_T^2}{\gamma^2}\;
  \frac{1}{\bbox{\Delta}^2+\frac{M_T^2}{\gamma}}.
\label{eqB22}
\end{eqnarray}

\vspace{0.4cm}  
  
\begin{eqnarray}
I^{\mu\nu\rho}_T  &=& -V^\mu_i V^\nu_i V^\rho_i
  \frac{i}{(4\pi)^2}\left(\frac{m}{\mu}\right)^3
  \int^1_0 d\alpha \frac{(1-\alpha)^3}{\alpha}
  \int^{1}_0 d\beta \frac{(1-\beta)^3}{\beta}\;
  \frac{2m\mu}{\bbox{\Delta}^2 + M^2_T} 
  \nonumber \\ [0.4cm]
&+&\left(g^{\mu\rho}V^\nu_i+g^{\nu\rho}V^\mu_i
  +g^{\mu\nu}V^\rho_i\right)
  \frac{i}{(4\pi)^2}\left(\frac{m}{\mu}\right)^2
  \int^1_0 d\alpha (1-\alpha)^2  \int^{1}_0 d\beta (1-\beta)
  \int_0^1 d\gamma \;
  \frac{M_T^2}{\gamma^2}\;
  \frac{1}{\bbox{\Delta}^2+\frac{M_T^2}{\gamma}}.
\label{eqB23}
\end{eqnarray}

\vspace{0.4cm}

\begin{eqnarray}
I^{\mu\nu\rho\sigma}_T  &=& -V^\mu_i V^\nu_i V^\rho_i V^\sigma_i
  \frac{i}{(4\pi)^2}\left(\frac{m}{\mu}\right)^4
  \int^1_0 d\alpha \frac{(1-\alpha)^4}{\alpha}
  \int^{1}_0 d\beta \frac{(1-\beta)^4}{\beta}\;
  \frac{2m\mu}{\bbox{\Delta}^2 + M^2_T} 
  \nonumber \\ [0.4cm]
&+&\left(g^{\mu\nu}V^\rho_iV^\sigma_i
  +g^{\nu\sigma}V^\mu_iV^\rho_i+g^{\rho\sigma}V^\mu_iV^\nu_i
  +g^{\mu\rho}V^\nu_iV^\sigma_i+g^{\nu\rho}V^\mu_iV^\sigma_i
  +g^{\mu\sigma}V^\rho_iV^\nu_i\right)
  \nonumber \\ [0.4cm]
&\times&  \frac{i}{(4\pi)^2}\left(\frac{m}{\mu}\right)^3
  \int^1_0 d\alpha (1-\alpha)^3  
  \int^{1}_0 d\beta (1-\beta)^2
  \int_0^1 d\gamma \;
  \frac{M_T^2}{\gamma^2}\;
  \frac{1}{\bbox{\Delta}^2+\frac{M_T^2}{\gamma}}
  \nonumber \\ [0.4cm]
&+& \frac{1}{2}\left(g^{\mu\nu}g^{\rho\sigma}+g^{\nu\rho}g^{\mu\sigma}
  +g^{\mu\rho}g^{\nu\sigma}\right)
  \frac{i}{(4\pi)^2}\left(\frac{m}{\mu}\right)
   \int^1_0 d\alpha\; \alpha(1-\alpha) \int^{1}_0 d\beta \;\beta M_T^2 
   \int_0^1 d\gamma \;
  \frac{M_T^2}{\gamma^2}\left(1-\frac{1}{\gamma}\right)
  \frac{1}{\bbox{\Delta}^2+\frac{M_T^2}{\gamma}}.
\label{eqB24}
\end{eqnarray}

\subsection{Integrals in configuration space:}

The configuration space integrals are obtained by Fourier transforming 
the results given above multiplied by $(-i)$ and by powers of the variable
$\theta=\left(\frac{t}{4\mu^2}-1\right)$. 
Recalling that $t=-\bbox{\Delta}^2$, we
have the general structure

\begin{equation}
S(r) = -i \frac{4\pi}{\mu^3} \int \frac{d^3 \bbox{\Delta}}{(2\pi)^3} \;
  e^{-i\protect\bbox{\Delta} \cdot \protect\bbox{ r}}\;
\left(\frac{-\bbox{\Delta}^2}{4\mu^2}-1
\right)^n
  \cdots \int_0^1 d\alpha \int_0^1 \cdots 
  \frac{1}{\bbox{\Delta}^2+M^2}.
\label{eqB25} 
\end{equation}

\twocolumn

Neglecting contact terms, we obtain

\begin{eqnarray}
&&S(r) = -i \frac{4\pi}{\mu^3} \int \frac{d^3 \bbox{\Delta}}{(2\pi)^3} \;
  e^{-i\protect\bbox{\Delta} \cdot \protect\bbox{ r}}\;
  \cdots \nonumber \\ [0.2cm]
&&\times\int_0^1 d\alpha \int_0^1 \cdots 
  \left(\frac{M^2}{4\mu^2}-1\right)^n\frac{1}{\bbox{\Delta}^2+M^2}
\label{eqB26}\\  [0.4cm]
&&= -i \cdots \int_0^1 d\alpha \int_0^1 \cdots 
  \left(\frac{M^2}{4\mu^2}-1\right)^n \,\frac{1}{\mu^2}\;\frac{e^{-Mr}}{\mu r}
\label{eqB27}\\ [0.4cm]
&&= -i \frac{1}{\mu r}\cdots \hat{\Theta}^n
  \int_0^1 d\alpha \int_0^1 \cdots 
  \frac{1}{\mu^2}\;e^{-Mr}.
\label{eqB28} 
\end{eqnarray}

\noindent 
where we use the short notation
\begin{equation}
\hat{\Theta}^n=\left(\frac{1}{4\mu^2}\frac{d^2}{dr^2}-1\right)^n .
\label{eqb29}
\end{equation}

In general, the integrals that enter Eqs.~(\ref{eq28}-\ref{eq37}) 
have at most
two free Lorentz indices, since the other ones are contracted with
powers of the vectors $V_i$. Therefore
integrals with one 
free Lorentz index are proportional to $V^\mu_i$ and those with
two  indices are proportional to either $V^\mu_iV^\nu_i$ or 
$g^{\mu\nu}$, motivating the definitions of 
Eqs.~(\ref{eq38},\ref{eq39}).  
In configuration space,
the terms originating from the representation of the logarithm 
have the form

\begin{equation}
S^{log}_n(M) = \int_0^1 dz \;
\frac{M^2}{\mu^2z^2}\;\left(1-\frac{1}{z}\right)^n
{e^{-\frac{M}{\sqrt z}r}},
\label{eqB30}
\end{equation}

These integrals can be evaluated explicitly, and we have

\begin{eqnarray}
S^{log}_0(M) &=& 2\frac{M^2}{\mu^2}
  \left[\frac{1}{Mr}+\frac{1}{(Mr)^2}\right] e^{-Mr},
\label{eqB31}\\ [0.4cm]
S^{log}_1(M) &=&  -4\frac{M^2}{\mu^2} 
  \left[\frac{1}{(Mr)^2}+\frac{3}{(Mr)^3}+\frac{3}{(Mr)^4}\right]
\nonumber \\
&\times& e^{-Mr},
\label{eqB32}\\ [0.4cm]
S^{log}_2(M) &=&  16\frac{M^2}{\mu^2} 
  \left[\frac{1}{(Mr)^3}+\frac{6}{(Mr)^4}
  +\frac{15}{(Mr)^5}+\frac{15}{(Mr)^6}\right] \nonumber \\ 
&\times& e^{-Mr}.
\label{eqB33}
\end{eqnarray}

For the bubble integrals this procedure yields the following 
results

\begin{eqnarray}
S_{B(0,n)} &=& \frac{1}{(4\pi)^2}\frac{1}{\mu r}
  \hat{\Theta}^n
  \int^1_0 d\alpha \, S^{log}_0(M_B), 
\label{Eq.34}\\ [0.3cm]
S_{B(0,n)}^{g} &=& \frac{1}{2}\frac{1}{(4\pi)^2}\frac{1}{\mu r}
  \hat{\Theta}^n
  \int^1_0 d\alpha \, S^{log}_1(M_B),
\label{Eq.35}\\[0.3cm]
S_{B(1,n)}^V &=& S_{B(0,n)}^{g},
\label{Eq.36}\\[0.3cm]
S_{B(2,n)} &=& S_{B(0,n)}^{g},
\label{Eq.37}\\[0.3cm]
S_{B(2,n)}^{g} &=& \frac{1}{8}\frac{1}{(4\pi)^2}\frac{1}{\mu r}
  \hat{\Theta}^n
  \int^1_0 d\alpha \, S^{log}_2(M_B), 
\label{Eq.38}\\[0.3cm]
S_{B(2,n)}^{VV} &=& 2 S_{B(2,n)}^{g},
\label{Eq.39}\\[0.3cm]
S_{B(3,n)}^{V} &=& 3 S_{B(2,n)}^{g},
\label{Eq.40}\\[0.3cm]
S_{B(4,n)} &=& 3 S_{B(2,n)}^{g}.
\label{Eq.41}
\end{eqnarray}

For the triangle integrals, we obtain

\begin{eqnarray}
S^V_{T(0,n)} &=& -\frac{1}{(4\pi)^2} \frac{1}{\mu r}\;
  2\left(\frac{m}{\mu}\right)^2\hat{\Theta}^n
  \int^1_0 d\alpha \frac{(1-\alpha)}{\alpha} \nonumber \\ [0.2cm]
&\times&  \int^{1}_{0} d\beta \; \frac{1-\beta}{\beta}\;
  e^{-M_Tr},
\label{eqB42}\\ [0.5cm]
S^{VV}_{T(0,n)} &=& -\frac{1}{(4\pi)^2}\frac{1}{\mu r}
  2\left(\frac{m}{\mu}\right)^3\hat{\Theta}^n
  \int^1_0 d\alpha \frac{(1-\alpha)^2}{\alpha}\nonumber \\ [0.2cm]
&\times&  \int^{1}_0 d\beta \frac{(1-\beta)^2}{\beta}\;
  e^{-M_Tr},
\label{eqB43}\\ [0.5cm]
S^{g}_{T(0,n)}  &=& \frac{1}{(4\pi)^2}\frac{1}{\mu r}
  (\frac{m}{\mu})\;\hat{\Theta}^n\int^1_0 d\alpha (1-\alpha)\nonumber \\ [0.2cm]
&\times&  \int^{1}_0 d\beta \; S_0^{log}(M_T)
\label{eqB44}\\ [0.5cm]
S^{VV}_{T(1,n)}  &=& -\frac{1}{(4\pi)^2}\frac{1}{\mu r}
  2\left(\frac{m}{\mu}\right)^4\hat{\Theta}^n
  \int^1_0 d\alpha \frac{(1-\alpha)^3}{\alpha}\nonumber \\ [0.2cm]
&\times&  \int^{1}_0 d\beta \frac{(1-\beta)^3}{\beta}\;
  e^{-M_Tr}+ 2S^{g}_{T(1,n)}, 
\label{eqB45}\\ [0.5cm]
S^{g}_{T(1,n)}  &=& \frac{1}{(4\pi)^2}\frac{1}{\mu r}
  \left(\frac{m}{\mu}\right)^2\hat{\Theta}^n
  \int^1_0 d\alpha (1-\alpha)^2\nonumber \\ [0.2cm]
&\times&  \int_0^1 d\beta (1-\beta) S_0^{log}(M_T),
\label{eqB46}\\ [0.5cm]
S^{VV}_{T(2,n)} & =& -\frac{1}{(4\pi)^2}\frac{1}{\mu r}
  2\left(\frac{m}{\mu}\right)^5\hat{\Theta}^n
  \int^1_0 d\alpha \frac{(1-\alpha)^4}{\alpha}\nonumber \\ [0.2cm]
&\times&  \int^{1}_0 d\beta \frac{(1-\beta)^4}{\beta}\;
  e^{-M_Tr}
  \nonumber\\ [0.2cm]
&+& 5 S^{g}_{T(2,n)} + 2 S^{g'}_{T(2,n)} 
\label{eqB47}\\[0.5cm]
S^{g}_{T(2,n)} &=& \frac{1}{(4\pi)^2}\frac{1}{\mu r}
  \left(\frac{m}{\mu}\right)^3\hat{\Theta}^n
  \int^1_0 d\alpha (1-\alpha)^3  \nonumber \\ [0.2cm]
&\times&  \int^{1}_0 d\beta (1-\beta)^2 \; S_0^{log}(M_T),
\label{eqB48}\\ [0.5cm]
S^{g'}_{T(2,n)} &=& \frac{1}{(4\pi)^2}\frac{1}{\mu r}\frac{1}{2}
  \frac{m}{\mu}\;\hat{\Theta}^n
  \int^1_0 d\alpha \alpha(1-\alpha)  \nonumber \\ [0.2cm]
&\times&   \int^{1}_0 d\beta \beta\; \frac{M_T^2}{\mu^2} \; 
  S_1^{log}(M_T).
\label{eqB49}\\ [0.5cm]
S^V_{T(1,n)}  &=& S^{VV}_{T(0,n)} + S^{g}_{T(0,n)}, 
\label{eqB50}\\[0.5cm]
S^V_{T(2,n)}  &=& S^{VV}_{T(1,n)} + S^{g}_{T(1,n)}, 
\label{eqB51}\\[0.5cm]
S^V_{T(3,n)}  &=& S^{VV}_{T(2,n)} + S^{g}_{T(2,n)} + S^{g'}_{T(2,n)}.
\label{eqB52}
\end{eqnarray}


\section{ANALYTICAL RESULTS FOR SOME INTEGRALS}

In this appendix we present analytic results for the asymptotic
bubble and triangle integrals in configuration 
space needed in this work.

The basic bubble integral is

\begin{equation}
S_{B(0,0)} = \frac{1}{(4\pi)^2}
 \; \frac{2}{\mu r}
  \int^1_0 d\alpha \frac{M_B^2}{\mu^2}
  \left[\frac{1}{M_Br}+\frac{1}{(M_Br)^2}\right] e^{-M_Br},
\label{eqC1}
\end{equation}

\noindent
where

\begin{equation}
M_B^2 = \frac{\mu^2}{\alpha (1-\alpha)}.
\label{Eq.C2}
\end{equation}

\noindent
Defining a new variable $t$ such that

\begin{equation}
\alpha = \frac{1}{2}+ \frac{t\sqrt{2+t^2}}{1+t^2},
\label{Eq.C3}
\end{equation}

\noindent
we have

\begin{eqnarray}
S_{B(0,0)}& = &\frac{1}{(4\pi)^2}
  2 \sqrt{2}\; \frac{1}{x}
  \int^{\infty}_0 dt \frac{1}{(1+t^2)^2\sqrt{1+\frac{t^2}{2}}}
 \nonumber \\ [0.4cm]
&\times&  \left[\frac{2(1+t^2)}{x}+\frac{1}{x^2}\right] e^{-2(1+t^2)x},
\label{eqC4}
\end{eqnarray}

\noindent
where $x=\mu r$. For large values of $x$, the integrand is very
peaked around $x\approx 0$ and hence we expand the functions in front
the exponential in a power series. Keeping the first three terms, 
we obtain our asymptotic expression

\begin{eqnarray}
S_{B(0,0)}^{asymp} &=& \frac{1}{(4\pi)^2}
  2 \sqrt{\pi}\; \frac{e^{-2x}}{x^{\frac{5}{2}}}
  \left(1+\frac{3}{16\,x}-\frac{15}{512\,x^2}+\cdots\right).
\label{eqC5}
\end{eqnarray}

For the triangle case, we have

\begin{eqnarray}
S^V_{T(0,0)} &= &-\frac{1}{(4\pi)^2} \frac{2}{\mu r}
  \int^1_0 d\alpha \frac{1}{\alpha(1-\alpha)}\nonumber \\ [0.4cm]
&\times&  \int^{\frac{m}{\mu}(1-\alpha)}_{0} ds \; \frac{s}{1-
\frac{\mu s}{m\,(1-\alpha)}}\;
  e^{-M_T\,r},
\label{eqC6}
\end{eqnarray}

\noindent
where we have used a new variable
$s\equiv (1-\beta)(1-\alpha)\frac{m}{\mu}$. The function 
$M_T^2$ is given by eq.(B21) and can be rewritten as

\begin{equation}
M_T^2 = M_B^2 \frac{\left[1-\frac{\mu}{m}s+s^2\right]}{1-
\frac{\mu s}{m\,(1-\alpha)}}.
\label{eqC7}
\end{equation}

\noindent
In the limit of $\frac{\mu}{m}\rightarrow 0$ we have

\begin{equation}
S^V_{T(0,0)} = -\frac{1}{(4\pi)^2} \frac{2}{\mu r}\;
  \int^1_0 d\alpha \frac{1}{\alpha(1-\alpha)}
  \int^{\infty}_{1} dy \; y e^{-M_B\,y\,r}\
\end{equation}

\noindent
where $y=\sqrt{1+s^2}$. Performing the \ $y$ \ integration and 
comparing it with Eq.~(\ref{eqC1}), we find

\begin{equation}
S^V_{T(0,0)} = - S_{B(0,0)}\;.
\label{eqC14}
\end{equation}


\end{document}